\documentclass[preprint,12pt]{elsarticle}

\usepackage{epsfig}
\usepackage{color}
\usepackage{graphicx}
\usepackage{subfigure}
\usepackage{natbib}

\usepackage{amssymb}

\journal{Annals of Nuclear Energy}

\begin{document}

\begin{frontmatter}



\title{3D Numerical Modeling of Liquid Metal Turbulent Flow in an Annular Linear Induction Pump}


\author[mephi]{K.I. Abdullina\corref{aki}}
\ead{kiabdullina@gmail.com}
\author[mephi]{S.V. Bogovalov}
\ead{svbogovalov@mephi.ru}
\author[ihte]{Yu.P. Zaikov}
\ead{dir@ihte.uran.ru}
\address[mephi]{National research nuclear university "MEPhI",\\ 
31, Kashirskoe shosse, Moscow, 115409, Russia}
\address[ihte]{Institute of high temperature electrochemistry RAN, Ural branch,\\
20, Akademicheskaya St., Ekaterinburg, 620137, Russia}

\begin{abstract}
3D numerical simulation of the liquid metal flow affected by electromagnetic field in the Annular Linear Induction Pump (ALIP) is performed using modified ANSYS package. ANSYS/EMAG  has been modified to take into account arbitrary velocity field at the calculation of the 
electromagnetic field and has been unified with CFX to provide integration of all the system of magnetohydrodynamic (MHD) equations defining dynamics of the conductive media in the variable electromagnetic field. The non-stationary problem is solved taking into account the influence of the metal flow on the electromagnetic field. The pump performance curve and the dependence of the pump efficiency on a flow rate are obtained.

\end{abstract}


\end{frontmatter}


\section{Introduction}
\label{Intro}

MHD ALIPs are the most perspective types of pumps for circulation of molten metals in fast breeder atomic reactors (FBRs). 
Although commonly used centrifugal pumps are more effective than existing ALIPs, the last ones have a number of advantages over the centrifugal pumps such as absence of moving parts and therefore high reliability, low noise and vibration level, simplicity of the flow rate regulation, easy maintenance and so on. Therefore, one of the basic problems in design of the ALIPs is the increase of their efficiency. It is expected that due to this the ALIPs will greatly simplify  design and improve industrial safety and efficiency of the FBRs. 

ALIPs have been widely investigated in  \cite{Kim12, KwantBoardman, Namba, YangKraus}. 
The characteristics of ALIPs were calculated in the model of MHD pump with narrow annular channel type. This type of pumps has been analyzed using an electrical equivalent-circuit method explored conventionally for induction machines. The relation between the developed pressure and the flow rate for various pump parameters has been obtained from the balance equation for the circuit \cite{Baranov}. However, such an approach can not be applied in some important cases. 

A lot of efforts have been devoted to develop ALIPs (see, for example, \cite{Andreev78, Andreev88, Kliman, Karasev, Rapin}) 
with large flow rate. The most significant problem that we are faced at the design of a large-scale pumps is the magnetohydrodynamic (MHD) instability. 
The instability arises when the Reynolds magnetic number exceeds one \cite{Gailitis}. Kirillov et al. \cite{Kirillov80} and Karasev et al. \cite{Karasev} 
experimentally exhibited that the magnetohydrodynamic instability is accompanied with a low frequency pulsation of the pressure and flow rate, azimuthal non-uniformity of the velocity and magnetic field, vibration of the pump and heat carrier loop, and fluctuation of the winding current and voltage. The conventional electrical methods does not help in this case. The numerical simulation  becomes the only reliable method of design of the large scale ALIPs. A lot of efforts were made to perform a computer modeling of the metal flow in the MHD pumps \cite{Obukhov, Galanin, NIIEFA1, Bergoug}. 
Nevertheless, all these attempts dealt with simplified models of the pumps, basically  two dimensional. The most advanced attempt was made in the work \cite{Obukhov}. 
However, the impact of the velocity of the metal on the electromagnetic field has not been incorporated into the code of this work. This approach did not allow to calculate the most important characteristics of the pump: P-Q characteristic and efficiency of the pump. 

Our main  objective  is  development of  a technology of computer design of ALIPs with high mass flow rate taking into account all the relevant physical processes. 
In this work we solve the particular problem of  development of a technology of numerical modeling of the metal flow in the channel of the pump based on the commercial numerical packages ANSYS/EMAG and ANSYS CFX temporally
neglecting the impact of the magnetic field on the turbulence. On this reason we limit our work by low magnetic Reynolds numbers. 

 ANSYS CFX provides calculation of hydrodynamics. We use ANSYS/EMAG to calculate transient electromagnetic fields in realistic 
physical conditions and in 3D geometry. ANSYS CFX by itself allows us to solve MHD problems for low values of the magnetic Reynolds numbers. However,
the original version of CFX does not allow us to model MHD flows in the necessary for our problem physical conditions and geometry. 
Therefore, we unified ANSYS CFX and ANSYS/EMAG into unique complex which allows us to model the flow of the metal in the conditions when the metal 
affects on the electromagnetic field and the electromagnetic field affects on the metal. 

The paper is organized as follows. In Section \ref{Modifications} we present the modifications of the ANSYS/EMAG and ANSYS CFX which were made for 
modeling of the MHD problems. Then in Section \ref{Model} we describe the computer model which was used to demonstrate the method. In section 
\ref{Results} we present the basic results.

\section{Modifications of CFX and ANSYS/EMAG}
\label{Modifications}

The basic requests to the numerical codes for solution of the industrial MHD  problems are: 
\begin{itemize}
\item ability to reproduce real geometry ;
\item ability to calculate eddy electromagnetic field in the conducting and in the surrounding media with different physical properties;
\item ability to solve the problem of non stationary MHD flow of the conducting media;
\end{itemize}
None of the numerical codes satisfy to these requests.  To minimize the work, we took the solution to modify one of the existing codes. The commercial package ANSYS is the most appropriate for this because it demands  smallest modifications.

The package ANSYS/CFX solves the hydrodynamical part of the problem while ANSYS/EMAG solves the electromagnetic part.
Due to our modifications the Lorentz force was included into the  system of hydrodynamical equations in ANSYS CFX and ANSYS/EMAG was modified 
to take in to account the impact of the velocity of the electrically conducting media on the electromagnetic field. The impact of the velocity on the Lorentz force was taken into account as well. 

In order to unify ANSYS CFX and ANSYS/EMAG into one software complex we developed a technology of synchronization of these packages and data exchange between them. 
 
\subsection{Basic equations}

The system of hydrodynamical equations includes the continuity equation
\begin{equation}
 {\partial \rho\over \partial t}+div\rho {\bf v}=0.
\label{cont}
\end{equation}
and the Navier-Stokes equation
\begin{equation}
\rho {\partial v_i\over\partial t}+\rho v_k{ \partial v_i\over \partial x_k}=-{\partial P\over \partial x_i}+ {\partial \tau_{ik}\over \partial x_k}+ [{\bf j\times \bf B}]_i.
\label{n-s}
\end{equation}
Here $\bf v$ is the velocity vector with  components $v_i$ and $v_k$,  $\rho$ - density, $P$ - pressure, $\tau_{ik}$ is the tensor of viscous stresses which includes molecular and eddy stresses. 
The last term in Eq.\ref{n-s} is the Lorentz force.

We used standard models of turbulence available in CFX. The impact of the magnetic field on the hydrodynamical turbulence is neglected in this work. The 
equations for turbulence are omitted here. They can be found in the manual for CFX \cite{ansyscfx}.  

Electromagnetic field  is calculated in ANSYS/EMAG in terms of vector potential $\bf A$ and electric potential $\phi$ integrated over time. The equations for $\bf A$ and $\phi$ are as follows
\begin{equation}\label{aeq}
\nabla\times{\nu}\nabla {\bf A}-\nabla \nu_{e}\nabla\cdot{\bf A}+{\sigma}({\partial{\bf A}\over \partial t}+\nabla{\partial \phi\over \partial t}
-[{\bf v\times[\nabla\times \bf A}]])=0.
\label{apot}
\end{equation}

\begin{equation}\label{veq}
\nabla\cdot(\sigma({\partial{\bf A}\over \partial t} +\nabla{\partial \phi\over \partial t}-[{\bf v}\times[\nabla\times{\bf A}]])=0.
\label{vpot}
\end{equation} 
 Here ${\nu}$  is the reversed magnetic permeability, ${\sigma}$ is the electric conductivity. The Coulomb gauge $div{\bf A}=0$ \cite{landau} 
is used for $\bf A$. The last terms in Eqs. (\ref{apot}) and (\ref{vpot}) appear due to motion of the electrically conducting media.

\subsection{Dependence of the electromagnetic field on the flow velocity.}

Simulation of the eddy electromagnetic field is available in ANSYS/EMAG  using finite element method \cite{galerkin}. Vector  $\bf A$ and electric $\phi$ potentials are specified at the nodes of the elements SOLID97 of ANSYS \cite{ansys}.  
Vector potential is interpolated in the finite element in the form 
$
A_i=\sum_{p}N_p({\bf r})A_i^p,
$
with summation over the nodes of the finite element. $N_p({\bf r})$ are the shape functions of the element. The electric potential has a  form $\phi=\sum_{p}N_p({\bf r})\phi^p.$
Galerkin method of discretization  gives the following discrete equations:
$
\hat C\dot u+\hat Ku=\hat J.
$
Here  vector 
$
u=\{A_x,A_y,A_z,\phi\}.
$
$\hat C$ is damping matrix which doesn't depend on the velocity. Stiffness matrix $\hat K$ has the following structure
\begin{equation}
{\hat K}=\left\{\begin{array}{cc}
K^{AA} & 0\\
K^{\phi A} & 0\\
\end{array}
\right\}.
\end{equation}

Matrix element  $K^{AA}$ has a form
\begin{equation}
K^{AA}=K^{AA}_0-\int [N_A]\sigma([{\bf v\times\nabla\times}[N_A]^T) d\Omega, 
\end{equation}
where integration $d\Omega$ is  performed over the volume of the finite element.   $K^{AA}_0$ is the matrix element corresponding to zero velocity of medium.  
Matrix element  $K^{\phi A}$ is fully defined by the velocity and has the following form
\begin{equation}
K^{\phi A}=-\int [\nabla N_A^T]^T\sigma([{\bf v\times\nabla\times}[N_A]^T) d\Omega.
\end{equation}
The correction of the  stiffness matrix has been performed by User Programmable Features \cite{ansys}.

\subsection{Verification  of the numerical code.}

\begin{figure}[htb]
\centering
\includegraphics[width=0.8\linewidth]{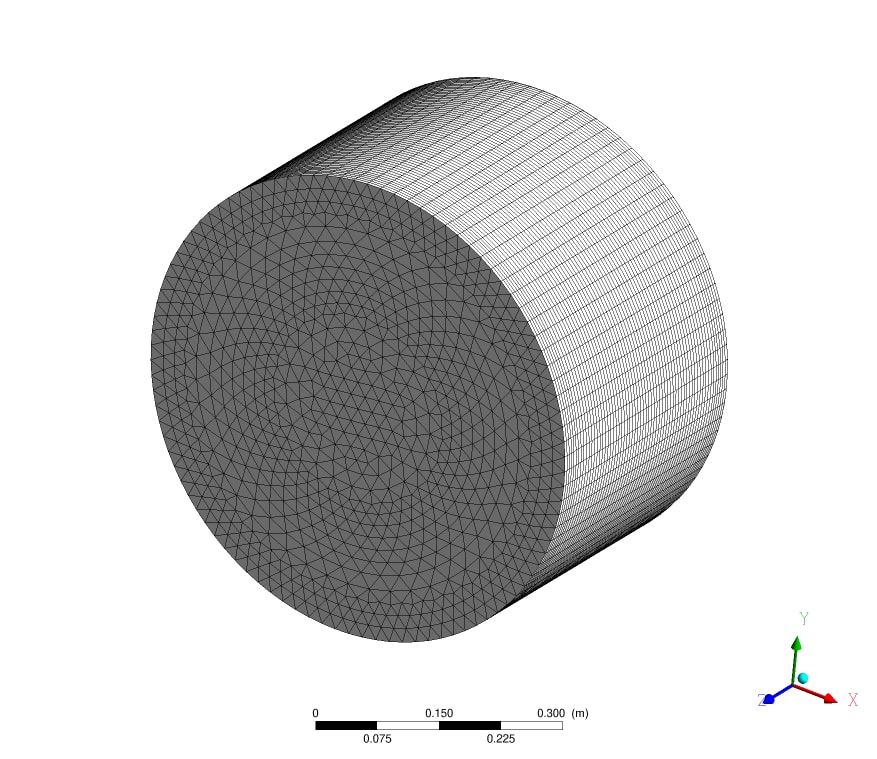}
\caption{Mesh of the test model}
\label{fig:dmesh}
\end{figure}

The objectives of the verification of the modified numerical code are the following:
\begin{itemize}
 \item To demonstrate that the code correctly reproduces the Lorentz force in ANSYS/CFX and the force correctly depends on the velocity of the conducting medium. 
 \item To make sure that the code is able to work with non hexagonal meshes.
 \item To make sure that the code can work with coarse meshes at magnetic Reynolds numbers exceeding 1.
\end{itemize}
 We choose rather simple model  consisting of a cylinder filled by a conducting medium. 
 Fig.~\ref{fig:dmesh} shows mesh of the model consisting of a cylinder with radius $R=30 ~\rm cm$ and length $L=36~\rm cm$.  
The cylinder is placed into  uniform magnetic field directed along the axis of the cylinder. The end cups of the cylinder are frozen into the magnetic field. One of the cups starts to rotate with a constant angular velocity $\omega=2\pi$ (frequency 1 Hz). The maximum speed at the edge of the end cup $V_{max}=\omega * R$ is about 1.9 m/s.  This velocity is chosen to consider the case with  the magnetic Reynolds number slightly exceeds 1. Indeed,   the electric conductivity is taken to be  $\sigma =3 \cdot 10^6~ \rm Ohm^{-1} \cdot m^{-1} $. The corresponding magnetic Reynolds number $R_m=vR\mu_0 \sigma$ is about 2 at $R=0.3 ~\rm m$, $v=1.9 ~\rm m/s$. $\mu_0$ is the magnetic permeability of vacuum. 

The dynamic molecular viscosity of the fluid is taken equal to zero to avoid excitation of rotation of the liquid by the viscous stresses. All the motion of the fluid occurs under the effect of the magnetic field. The motion is laminar.   

After beginning of rotation an Alfvenic wave propagates along the cylinder. The velocity of propagation equals to \cite{landau}
\begin{equation}
 v_A={B_0\over \sqrt{\mu_0 \rho}},
 \label{alfven}
\end{equation}
where $B_0$ is the initial magnetic field,  and $\rho$ is the density of the fluid.  The key parameters which are used for the verification are the velocity of propagation of the wave and velocity of rotation of the fluid.  These parameters crucially depend on the correctness of calculation of the Lorentz force in CFX and its dependence on the velocity of medium. 

\begin{figure}[!htb]
\centering
\includegraphics[width=0.8\linewidth]{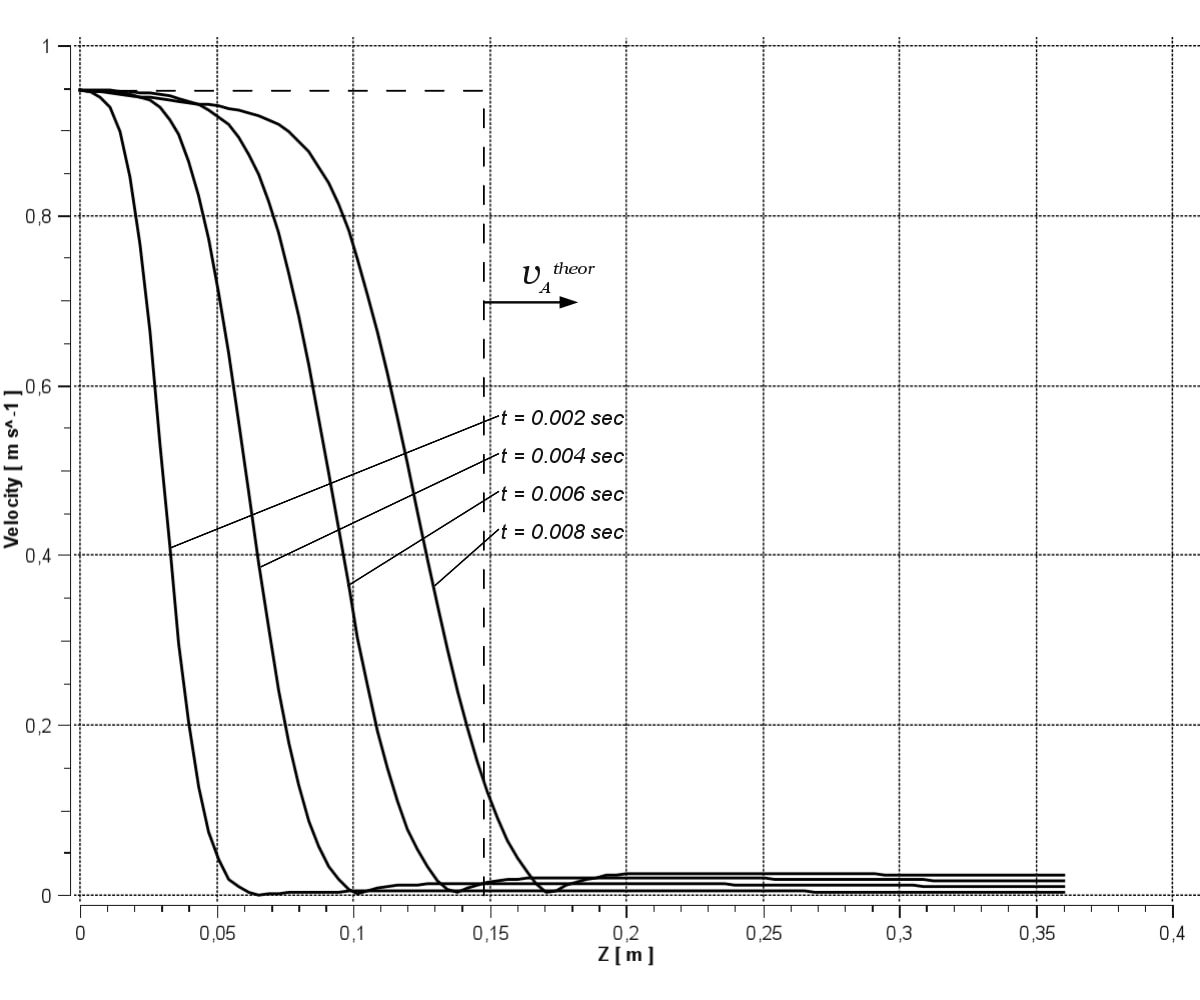}
\caption{Dependence of velocity on $z$ for different moments of time at $r=15 ~\rm cm$. Dashed line shows theoretical distribution 
of the velocity in the limit of infinite electric conductivity of the medium. }
\label{fig:dwave}
\end{figure}

Theoretically, an Alfvenic wave of rotation has to propagate along the cylinder with the velocity (\ref{alfven}) in the limit of infinite electric conductivity of the medium.
The velocity of the medium has to be equal to the  velocity of rotation of the end cup \cite{akhiezer}. In other words the distribution of the velocity along z coordinate should be a step function shown by dashed line in Fig. \ref{fig:dwave}. 

Dependence of the velocity along z coordinate for different moments of time is presented in Fig.~\ref{fig:dwave}. Dashed line shows the theoretical shape of the velocity distribution of propagation of the wave at the moment $t=0.008 ~\rm sec$ from the beginning of rotation. 
The velocity of propagation $v_A^{cal} \sim 17 ~\rm m/s$ is close to the theoretical value $v_A^{theor} =18.6 ~\rm m/s$. The difference less than 10\% is explained by the coarse mesh in $r$ direction.
The shape of the distribution of the velocity along $z$ coordinate differs from the step function on evident reasons. The step function is obtained in the limit of ideal  magnetohydrodynamics when the electric conductivity of the medium goes to infinity. The calculations were performed at the finite
conductivity equal to $\sigma =3 \cdot 10^6~ \rm Ohm^{-1} \cdot m^{-1}$.   The widening of the step function is defined by the expression $\Delta z = 2\sqrt{t/(\sigma\mu_0)}$ and equals  $\sim 9 ~\rm cm $ at the selected parameters. This fully explains the difference between the theoretical prediction and  calculated distribution of the azimuthal velocity of the medium along  $z$. 

\begin{figure}[!htb]
\centering
\includegraphics[width=0.7\linewidth]{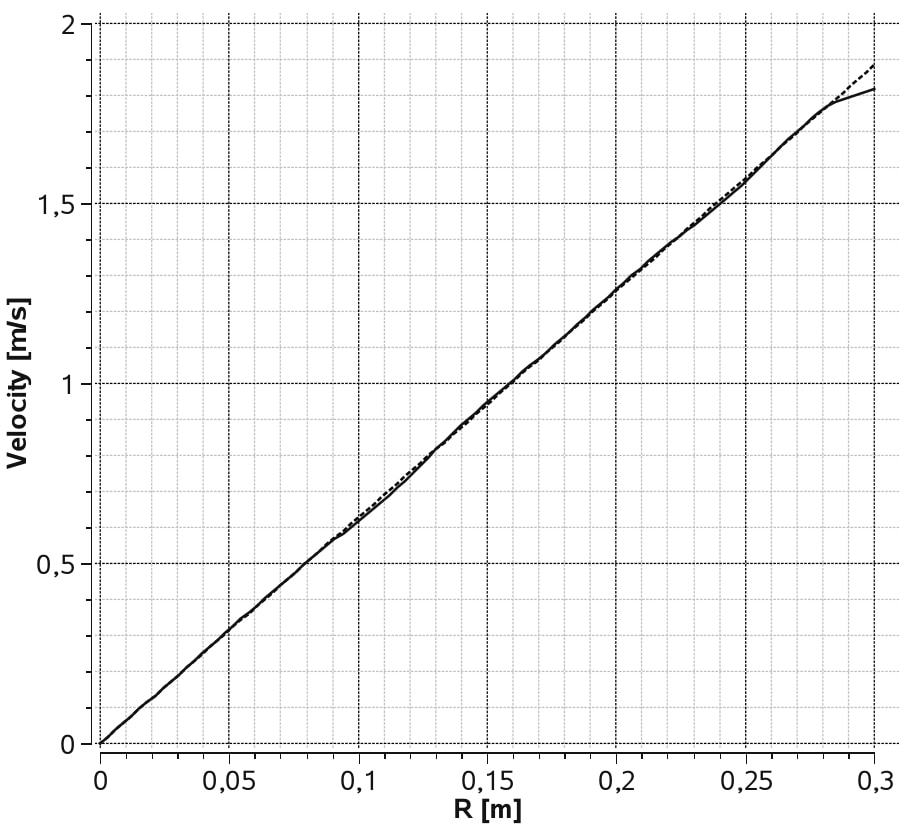}
\caption{Dependence of the velocity on radius. Dashed line corresponds to the rigid body rotation}
\label{fig:dvelg}
\end{figure}

The velocity distribution along the radius of the cylinder well agree with the theoretical distribution for rigid rotation of 
the liquid (see Fig.~\ref{fig:dvelg}).

The test problem has been solved at the moderate magnetic Reynolds number $R_m \sim 2$. Apparently this is sufficient for modeling of the MHD pumps with high mass flux corresponding to the magnetic Reynolds number of the order 1.


\section{Model description}
\label{Model}

\begin{figure}[!htbp]
\centering
\includegraphics[width=1\linewidth]{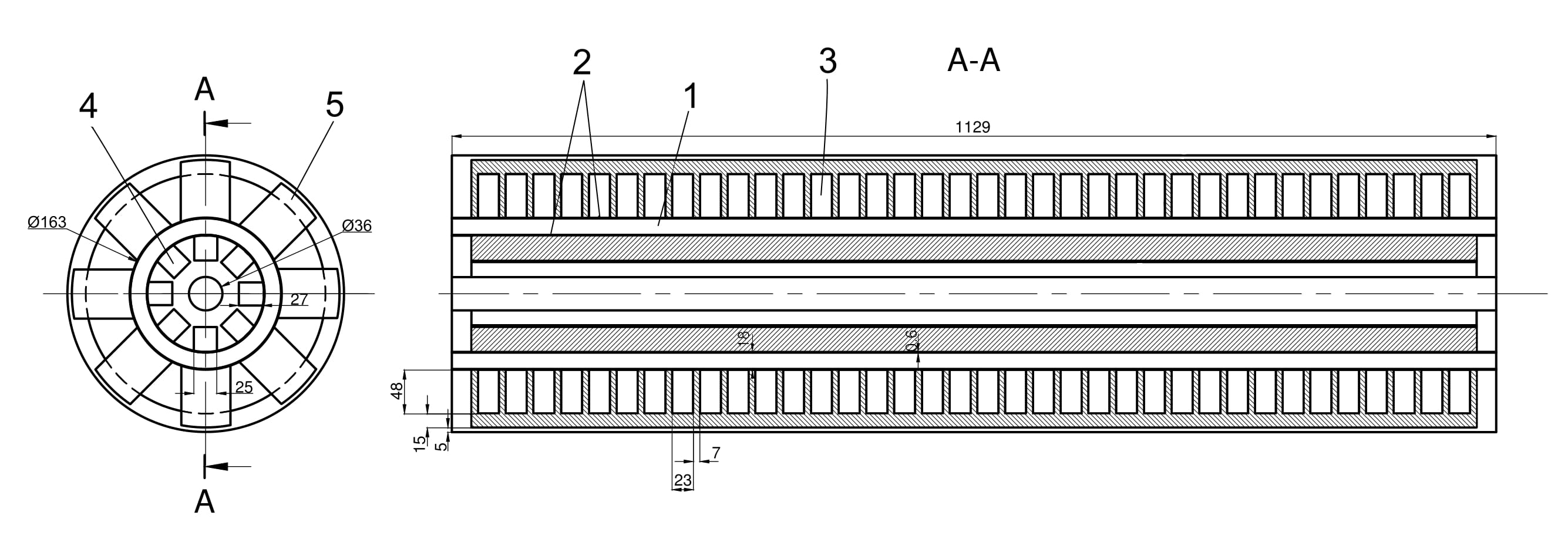}
\caption{Cross-sectional view of the ALIP. 1 -- working channel filled by the liquid metal, 2 -- thin shells forming a channel gap, 3 -- winding coils, 4 -- inner core, 5 -- outer core}
\label{fig:scheme}
\end{figure}

The main parts of the pump are the inner and the outer cores with high magnetic permeability, three-phase winding coils and a narrow annular channel gap between shells of the inner and the outer cores. The design of the modeled ALIP is based on \cite{patent}. Fig.~\ref{fig:scheme} shows the 
geometry of the computational domain.

\begin{figure}[htbp]
\centering
\includegraphics[width=0.8\linewidth]{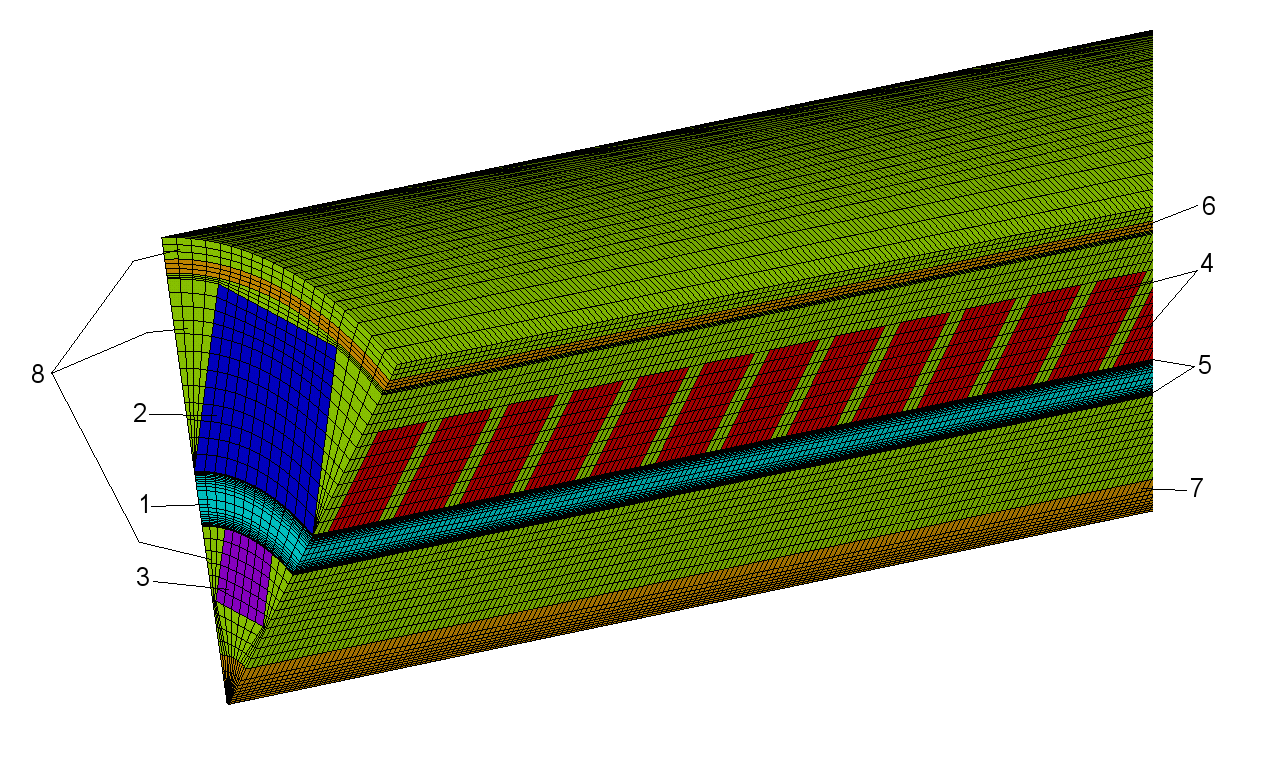}
\caption{Mesh fragment of the MHD pump computation domain. 1~-- liquid metal, 2~-- outer core, 3~-- inner core, 4~-- coils of inductor, 5~-- shells forming annular channel gap, 6~-- housing of the pump, 7~-- shift, 8~-- air}
\label{fig:model}
\end{figure}

The mesh has been generated by the mesh generator ICEM CFD. Hexagonal elements SOLID97 enabling calculation of the vector and scalar 
potentials on the nodes have been used. The outer boundaries of the computational domain are placed away from the pump. The pump is surrounded by air (green elements (8) in Fig.~\ref{fig:model}) and the outer layer elements extend the solution to infinity using a standard ANSYS method. 

The computational domain (Fig. \ref{fig:model}) includes only a $45^{\circ}$-sector of the full model of the pump, due to the cyclic symmetry of the pump geometry and the excited electromagnetic fields. This  allows us to reduce the computation time and relax requirements to the hardware. 

Constant magnetic permeability of the ferromagnetic cores instead of the nonlinear permeability has been used. 
This simplification can be applied as found in \cite {Abdullina}. The value of constant magnetic permeability $\mu=1289$ has been obtained as an average value calculated by nonlinear analysis.

The cores are composed of the sheets of electrical steel. The eddy electric currents don't flow in azimuthal direction. The pump housing and the channel shells are made of conductive materials as well.

Pb-Bi alloy is taken as the pumped medium. The physical characteristics of the alloy needed for the calculations are given in Table~\ref{tab1}.
\begin{table}[!htbp]
\centering
\caption{\label{tab1}Pb-Bi alloy characteristics}
\begin{tabular}{l|l}
\hline
Characteristic & Value  \\
\hline
Density $\rho, {\rm kg/m^3}$ & 10360 \\
Specific electric conductivity $\sigma, {\rm 1/Ohm \cdotp m}$ & $0.847 \cdotp 10^6$\\
Dynamic viscosity $\eta, {\rm Pa\cdotp s}$ & $1.94 \cdotp 10^{-3}$\\
\hline
\end{tabular}
\end{table}

\begin{figure}[htbp]
\centering
\includegraphics[width=1\linewidth]{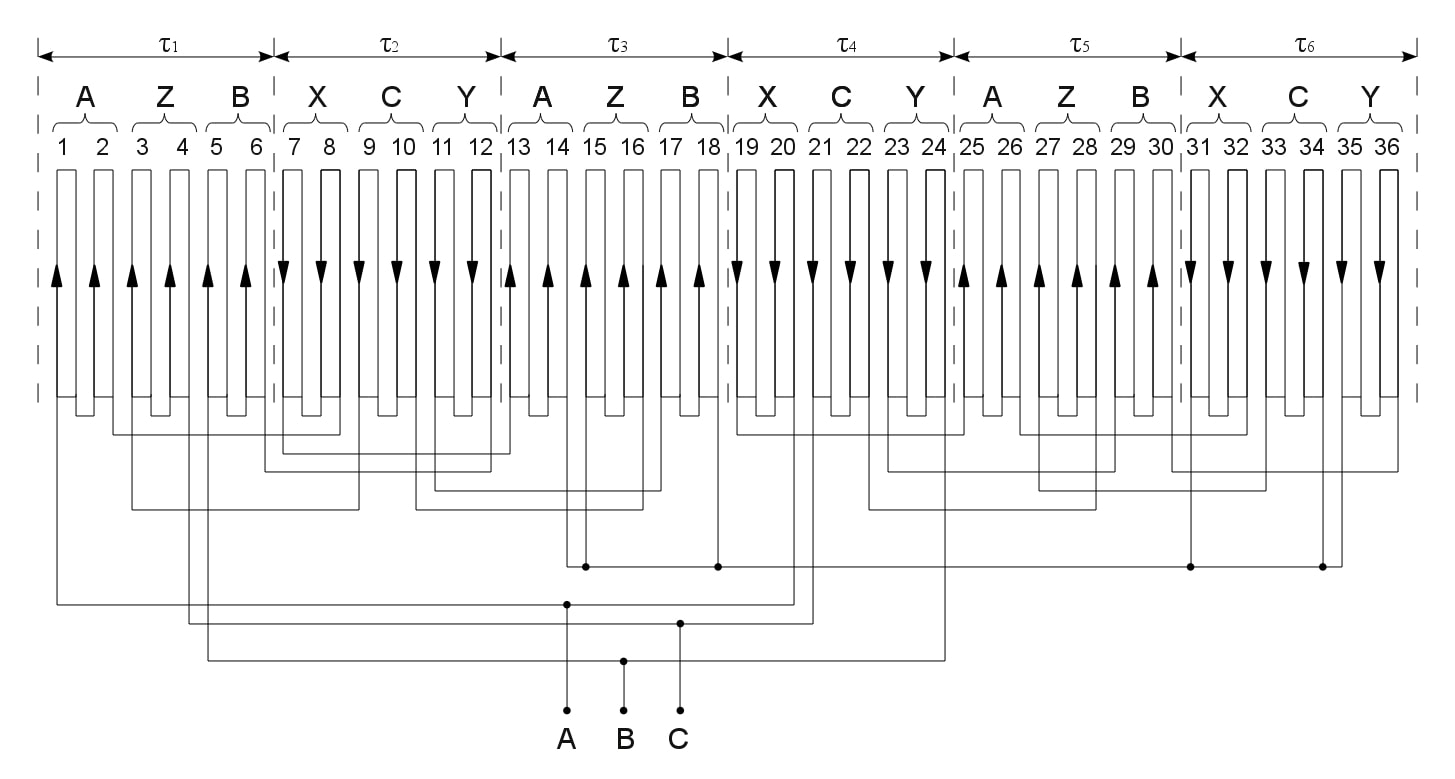}
\caption{Wiring diagram of the inductor. A, B, C -- phases; $\tau$ -- pole pitch}
\label{fig:connect}
\end{figure}

Transient analysis have been performed to calculate alternating electromagnetic field and velocity distribution of the metal. 
Flow of the metal is considered to be turbulent and SST turbulence model is used. The time step of the transient analysis equals to
$1.25 \cdotp10^{-3}$ seconds. The three phase power supply has frequency  50 Hz. Current density $\bf j$ 
has been specified in the coils.  Wiring diagram is shown in Fig.~\ref{fig:connect}.

\section{Results}
\label{Results}
 \subsection{Performance curve}

\begin{figure}[htbp]
\centering
\includegraphics[width=0.9\linewidth]{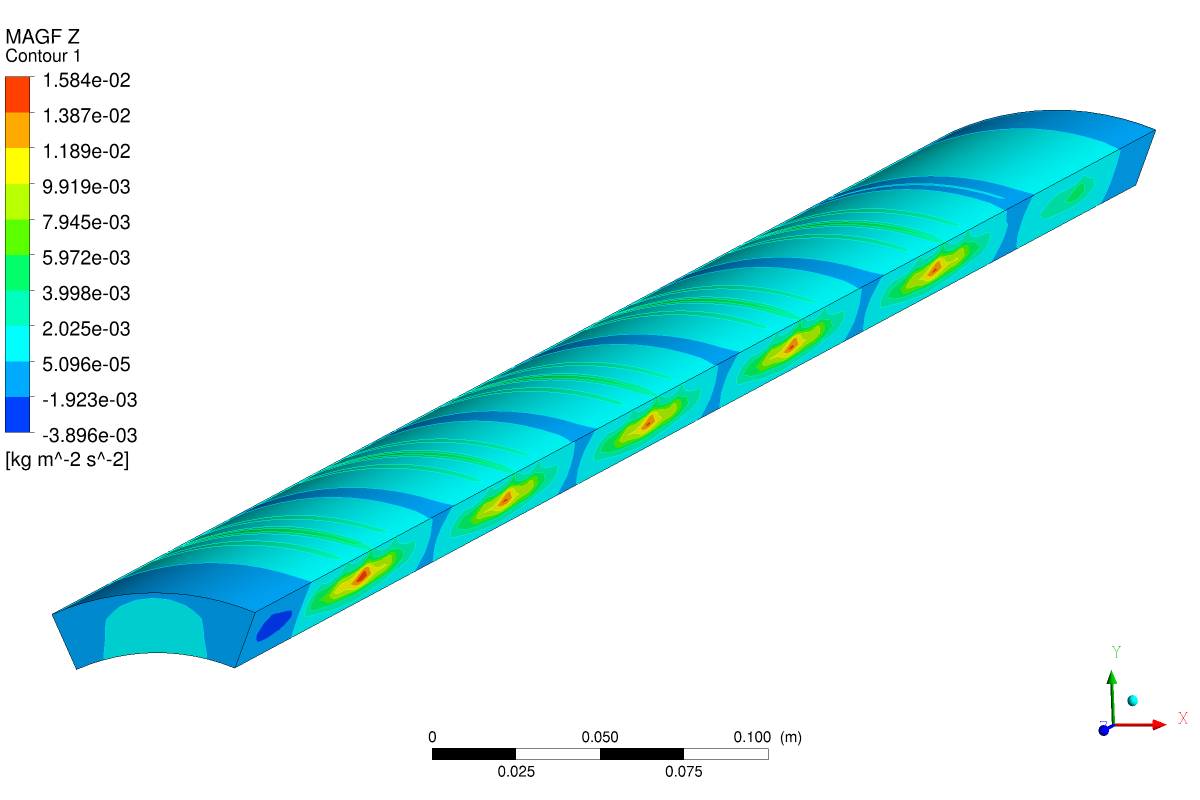}
\caption{Axial component of the Lorenz force $F_z$ in the channel at some time moment}
\label{fig:force}
\end{figure}

\begin{figure}[htbp]
\centering
\includegraphics[width=0.9\linewidth]{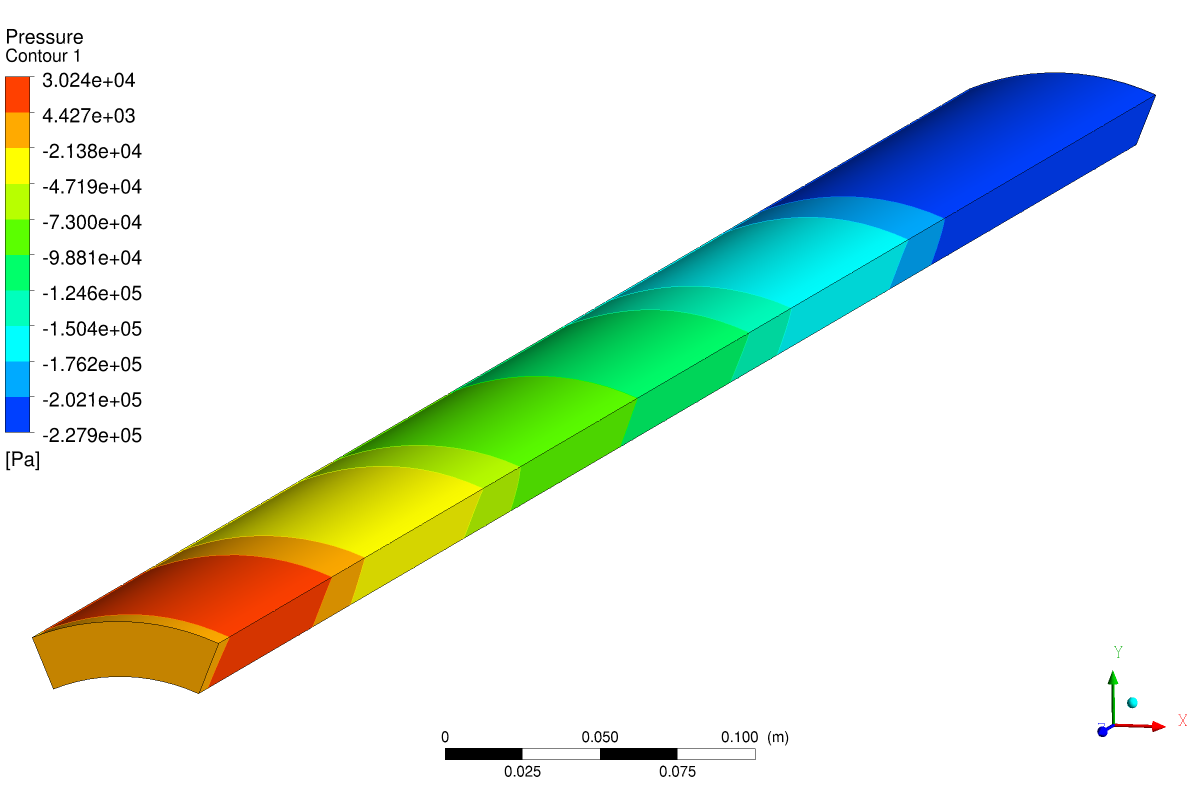}
\caption{Distribution of relative pressure compared to outlet}
\label{fig:pres}
\end{figure}

A series of calculations has been performed for different flow rates of the metal. All the problem is solved as non stationary. Fig.~\ref{fig:force} shows distribution of the Lorenz force in the channel at flow rate $\rm Q=0.046 \; m^3/s$ in some time moment. The force is basically periodic. The periodicity is violated only at the ends of the channel (the end effects). The force becomes directed opposite to the direction of the flow at the end closest to the reader (outlet).  The force is positive but also differs from the forces in the central part of the channel at the far end (inlet) .  The force makes the liquid metal to flow, and produces the  pressure drop in the pump (Fig.~\ref{fig:pres}). The pressure presented in the figure is relative pressure with zero specified at the outlet. Absolute pressure for the incompressible liquid has no physical meaning. Only pressure drop has physical sense.
Therefore, for the correct understanding of this figure it necessary to keep in mind that the absolute pressure equals to 1 atm at the inlet.

\begin{figure}[htbp]
\centering
\includegraphics[width=0.8\linewidth]{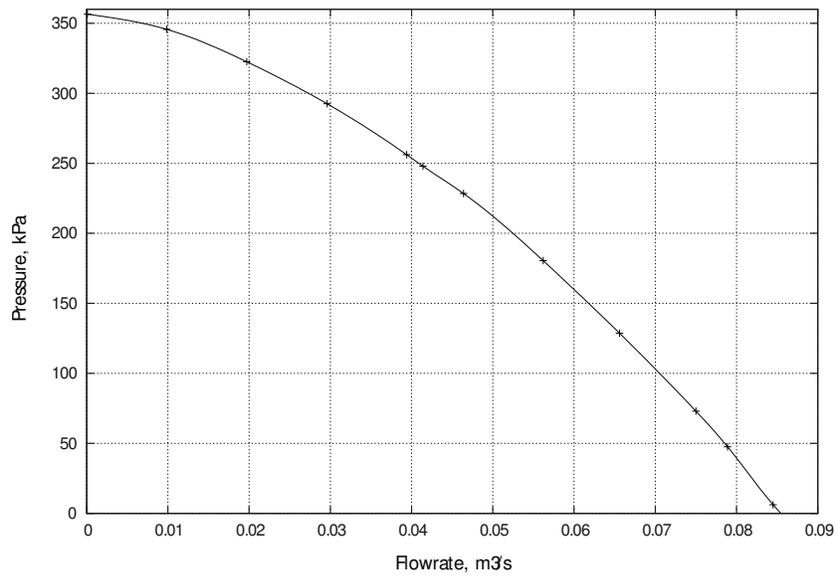}
\caption{MHD pump performance curve}
\label{fig:curve}
\end{figure}

The main result of the calculations is MHD pump performance curve shown in Fig.~\ref{fig:curve}. The curve has a conventional parabolic shape, 
which is the result of the  pressure drop produced by turbulent flow and the impact of velocity field of the metal on the Lorenz force. 

\subsection{Efficiency of the pump}
The power consumption $W$ (per one coil) equals to
\begin{equation}
 W(t)=\int{\bf E}(t){\bf j}(t)dV.
\label{pow}
\end{equation}

Electric field ${\bf E}$ can be expressed through  vector ${\bf A}$ and scalar $\phi_e$ potentials as
\begin{equation}\label{efield}
 {\bf E}=-\frac{\partial {\bf A}}{\partial t} - \nabla\phi_e.
\end{equation}
Our estimates show that contribution of the electrostatic field produced by Ohmic resistivity of the coils is negligible. Only eddy electric field in Eq.\ref{pow} is taken into account at the calculation of the power consumption. 

The current density in the inductor has only azimuthal component $j=j_\phi$. The electric current varies sinusoidally $I(t)=I_0\sin\omega t$. The 
average power consumption of the inductor of 36 coils is defined as follows

\begin{equation}
 W_0=36 I U{\cos\phi},
\end{equation}
where $I$ and $U$ are the working current and voltage drop in one coil. The voltage drop is defined as $U=n\int E dl$, where $n$ is amount of turns of 
the wire in the coil,  and integration is performed over the turn of wire in the coil.

\begin{figure}[htbp]
\centering
\includegraphics[width=0.9\linewidth]{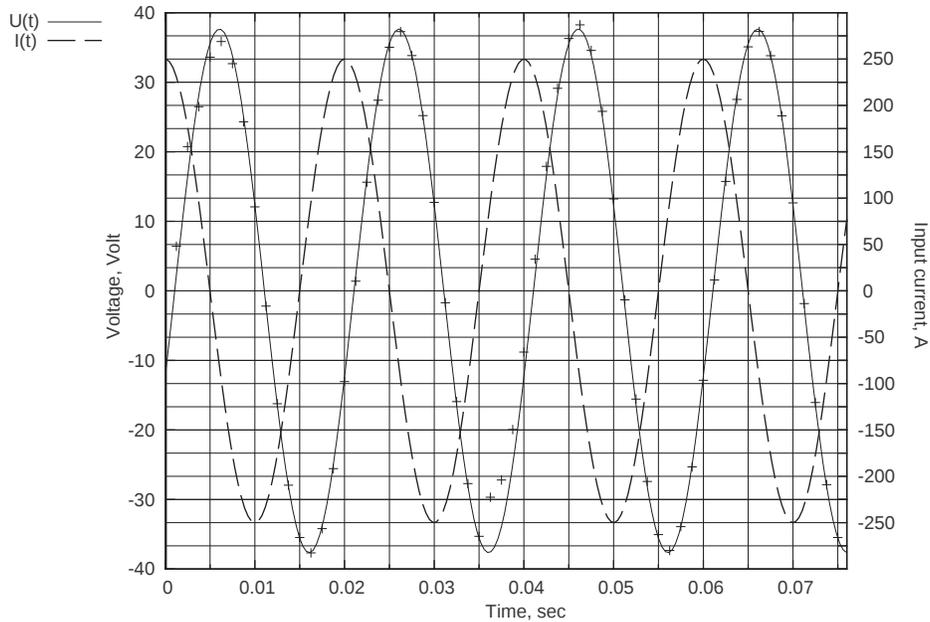}
\caption{Time dependence of the voltage drop and  current in one of the coils. Solid line represents sine function fitting the results, dashed line shows the input current}
\label{fig:sin}
\end{figure}

The phase shift $\phi$ between the voltage and current has been obtained graphically (see Fig.\ref{fig:sin}).
The power factor $\cos\phi$ has been obtained with an error defined by the length of the data series. The error $\Delta\phi$ 
is inversely proportional to the length of the data series. The larger the length the less the error. Because of limitations of 
the disk space (ANSYS/EMAG creates huge files of results) we  saved the data of the limited length.  Therefore, all the data containing the information about the angle $\phi$ are plotted with the errors 
originating from accuracy of the calculation procedure, not from any experimental data.

\begin{figure}[htbp]
\centering
\includegraphics[width=0.8\linewidth]{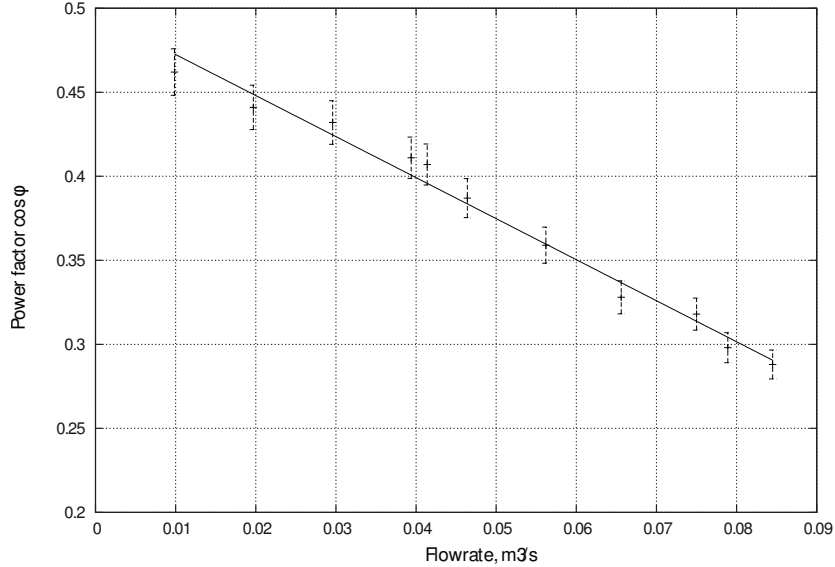}
\caption{Dependence of the power factor $\cos\phi$ on a flow rate}
\label{fig:cos}
\end{figure}

\begin{figure}[htbp]
\centering
\includegraphics[width=0.8\linewidth]{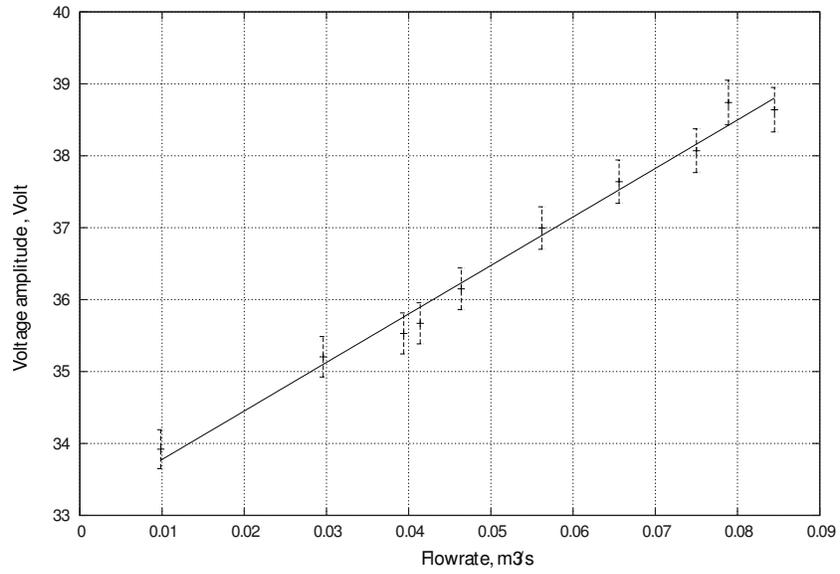}
\caption{Dependence of the voltage drop amplitude of a coil on a flow rate}
\label{fig:volt}
\end{figure}

Dependence of the power factor $\cos\phi$ and voltage drop amplitude of a coil on the flow rate is shown in Fig.~\ref{fig:cos} and Fig.~\ref{fig:volt} respectively. 
These dependencies are linear within the errors. 

\begin{figure}[!htbp]
\centering
\includegraphics[width=0.8\linewidth]{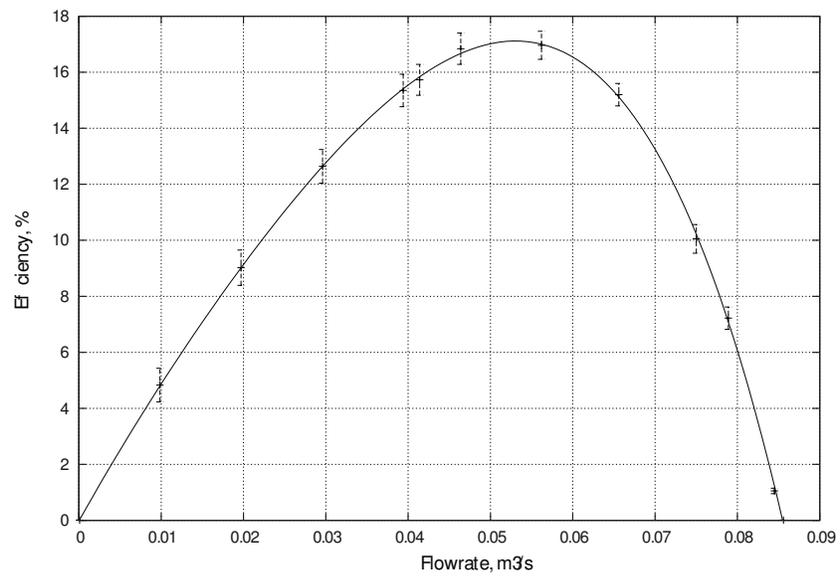}
\caption{Dependence of the pump efficiency on a flow rate}
\label{fig:kpd}
\end{figure}
Resulting dependence of the pump efficiency on a flow rate is shown in Fig.~\ref{fig:kpd}. It is also calculated with an error up to 4\%
The efficiency of this design of MHD pump doesn't exceed 17\% (see Fig.~\ref{fig:kpd}).

\section{Discussion}
\label{Conclusion}

The basic result of this work is  development of a computer technology  for numerical study of 3-D flows in ALIPs at different regimes of flow, provided that the magnetic Reynolds number does not strongly exceed value  of the order of 1. 
The proposed computer complex is developed on the basis of the ANSYS products ANSYS/EMAG and ANSYS/CFX.  Unlike the conventional electrotechnical calculations and 2-D numerical calculations performed before,  the developed complex gives us the following advantages:
\begin{itemize}
 \item the pump performance and efficiency curves can be calculated in all regimes of the work of the pump;
 \item detailed geometry and physical characteristics of the materials can be taken into account at the calculations;
 \item non stationary  regime of the MHD flow in the pump can be investigated by this computer complex.
\end{itemize}

In the work we have demonstrated verification of the code on the nonstationary test problem. The code has been  applied  to the calculation of the characteristics of real MHD pump which can be obtained only experimentally up to now. Unfortunately, we can not give direct comparison with experiments. The experimental data presented in the literature never gives details about the design of the pumps, while the publications with open information about design of the pump never give the the experimental measurements. Nevertheless, for the pump described in this work we achieved the agreement with experiment in the limit of 5\% for the P-Q curve at the maximum of efficiency of the pump. 

In this work we did not take into account the impact of the magnetic field on the hydrodynamic turbulence. For the specific pump discussed here  this is apparently not important. Indeed, the size of the gap between the coaxial cylinders where the liquid metal flows $\sim 2~ \rm cm$.
Even at the largest possible flow rates $0.085~ \rm m^3/s$ the velocity of the metal equals $\sim 12~ \rm m/s$. In these conditions the magnetic Reynolds number does not exceeds 0.24. Pretty well  below 1.  Therefore, it is not surprising that the impact of the magnetic field on the turbulence is negligible in the pump.   Consideration of the pumps with larger mass flow rates will certainly request modification of the turbulence models. But this is separate important problem. It can be 
solved in large work with detailed comparison of the calculations with experiments.

\section*{Acknowledgments} 
The work of S.Bogovalov and K.Abdullina has been supported by grant no.16-12-10443  of the Russian scientific fund. The work of Yu. Zaikov
has been supported by the grant no.14.607.21.0146 (ID of the project RFMEFI60716X0146) of Ministry of education and science of Russia.



\bibliographystyle{elsarticle-num.bst} 


\end{document}